\documentclass[11pt]{article}
\pdfoutput=1
\usepackage{epsfig}
\usepackage{latexsym}
\usepackage{amssymb}
\usepackage{amsmath}
\usepackage[hypertex]{hyperref}
\usepackage{graphicx}
\usepackage{axodraw}

\begin{document}

\pagestyle{empty}



\qquad\\
\qquad\\
\qquad\\

\begin{center}
\mbox{\bf\LARGE  Signals of New Physics} \\ [11pt]
\mbox{\bf\LARGE in the Underlying Event}

\vspace*{1.3cm}
{\large Roni Harnik$^{1,2}$, Tommer Wizansky$^2$} \\
\vspace*{1.3cm}

\mbox{$^1$\textit{Stanford Institute of Theoretical Physics, 
Stanford University, Stanford, CA 94309}} \\[2mm]
\mbox{$^2$\textit{Stanford Linear Accelerator Center, 
Stanford University, Stanford, CA 94309}} \\

\vspace*{1cm}

\texttt{roni@slac.stanford.edu},~ 
\texttt{twizansk@slac.stanford.edu}

\end{center}

\vspace*{0.2cm}

\begin{abstract}

LHC searches for new physics focus on  combinations of hard physics objects. In this work 
we propose a qualitatively different \emph{soft} signal for new physics at the LHC -- the ``anomalous underlying event''. Every hard LHC event will be accompanied by a soft underlying event due to QCD and pile-up effects. Though it is often used for QCD and monte carlo studies, here we propose the incorporation of an underlying event analysis in some searches for new physics. An excess of anomalous underlying events may be a smoking-gun signal for particular new physics scenarios such as ``quirks'' or ``hidden valleys'' in which large amounts of energy may be emitted by a large multiplicity of soft particles. 
We discuss possible search strategies for such soft diffuse signals in the tracking system and calorimetry of the LHC experiments. We present a detailed study of the calorimetric signal in a concrete example, a simple quirk model motivated by folded supersymmetry. In these models the production and radiative decay of highly excited quirk bound states leads to an ``antenna pattern'' of soft unclustered energy. Using a dedicated simulation of a toy detector and a ``CMB-like'' multipole analysis we compare the signal to the expected backgrounds.



\end{abstract} 

\newpage
\pagestyle{plain}

\section{ A New Signal: the Anomalous Underlying Event}\label{intro}

The LHC is about to probe the physics of the TeV scale. It holds great hope to discover new physics beyond the standard model and shed light on electroweak symmetry breaking and dark matter. 
The LHC detectors are designed to accurately detect several ``physics objects'' such as hard leptons, photons, jets, all of which are prompt, in addition to missing transverse energy and so on. Searching for new physics typically involves looking for an excess of events with a particular combination of such physics objects after an appropriate set of cuts is employed. Some of the conceptually simple searches include looking for an excess of events with leptons and/or jets and missing energy in models which include a WIMP dark matter candidate, or resonance peaks in the kinematic distributions of leptons in models with a new $Z'$ gauge boson. The standard list of physics objects is also used for triggering.

In addition to the standard set of physics objects, one can think of models which give rise to very non-standard objects. What is the value in these unconventional signals?
Consider, for example, highly displaced vertices which were highlighted as potential signals for Hidden Valley models~\cite{hv1, hv2, hv3}. Discovery of highly displaced vertices can teach us that a new neighbor sector exists alongside the standard model. Studying the details of these events may teach us about the properties of the new sector. 
Another example is that of a stopped stable gluino that decays out of time in the hadronic 
calorimeter~\cite{stopping}.
Such events can allow us to measure the lifetime of the gluino in split SUSY~\cite{split} and infer valuable knowledge about high energy parameters such a the SUSY breaking scale.
Even though such signals are not what the detectors were designed to discover, non-standard analyses have shown that they may be discovered with a reasonable efficiency. If discovered, these signals may give us new and complementary information about the new physics at the TeV scale.

In this paper we propose a new observable physics object at colliders -- an anomalous underlying event. 
The underlying event (UE) is the often overlooked part of every LHC event.
It is defined as everything in the event which is not the outgoing hard jets or leptons. Because of its omnipresence, the UE will be carefully studied and characterized for a variety of hard final states in terms of distributions of soft tracks and unclustered energy. Characterizing the UE will be a valuable tool for tuning Monte Carlo event generators. 
An \emph{anomalous} underlying event is one which contains a distinctively uncharacteristic distribution of soft tracks and diffuse energy when compared to a characteristic UE  with a similar hard final state. Like any candidate signal, anomalous underlying events may and will be produced by SM backgrounds. Discovery of new physics with anomalous UE's requires a statistically significant number of anomalous UE's when compared to the SM expectation.  Claiming such a discovery with confidence  will require extracting the characteristics and of the typical UE and their variations from other LHC data.

\begin{figure}[t!]
 \centering
  \includegraphics[height=7cm]{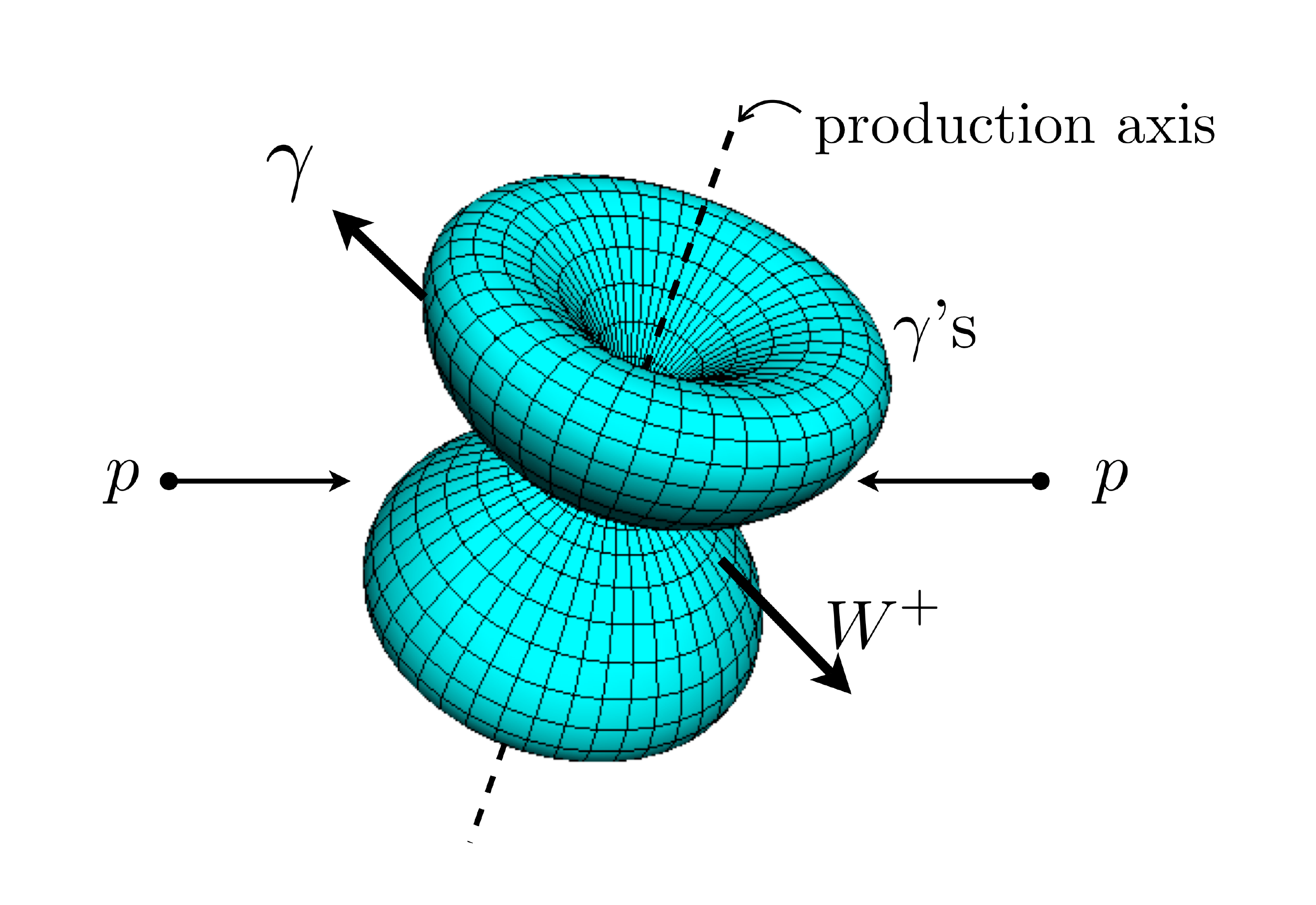}
   \caption{A schematic cartoon of the initial and final states of an LHC event with squirk production via an s-channel $W^\pm$. The two protons are incoming along the horizontal axis. The squirks are produced and oscillate along the dashed axis.
The final state includes an antenna pattern of soft photons (two cone like shapes aligned with the squirk production axis) and a pair of hard annihilation products, $W\gamma$ in this case. The search strategy will first involve discovering a resonance in $W\gamma$ and then searching for signals of patterns of soft photons in the candidate signal events.   
}
  \label{fig-cartoon}
\end{figure}

The signal and its search strategy are best explained in a concrete example.
We will thus briefly explain the model and signal which are the focus of this paper and sketch the search strategy for this example. We will defer both details and general lessons to subsequent sections. 
In this paper we will consider a new physics scenario in which some new particles with a mass of several hundred GeV are pair produced at the LHC. The new particles are charged under a new strong force and are thus confined to a single highly excited mesonic bound state. Due to the 
``quirky''~\cite{quirks} nature of the new dynamics the excitation energy may easily exceed several hundreds of GeV. After production, the excited bound state will emit soft radiation, and decay to the ground state, emitting many quanta. Some of these quanta will be \emph{soft} photons which are emitted in a particular angular distribution. The ground state will then annihilate into a \emph{hard} final state, for example a hard $W^\pm$ and a hard photon. The invariant mass of the $W$+photon system reconstructs to the mass of the ground state meson (again, at several hundred GeV). All of the processes discussed above are prompt on collider time scales.
A cartoon initial and final states of these events are depicted in Figure~\ref{fig-cartoon}. 

\begin{figure}[t!]
 \centering
  \includegraphics[height=15cm]{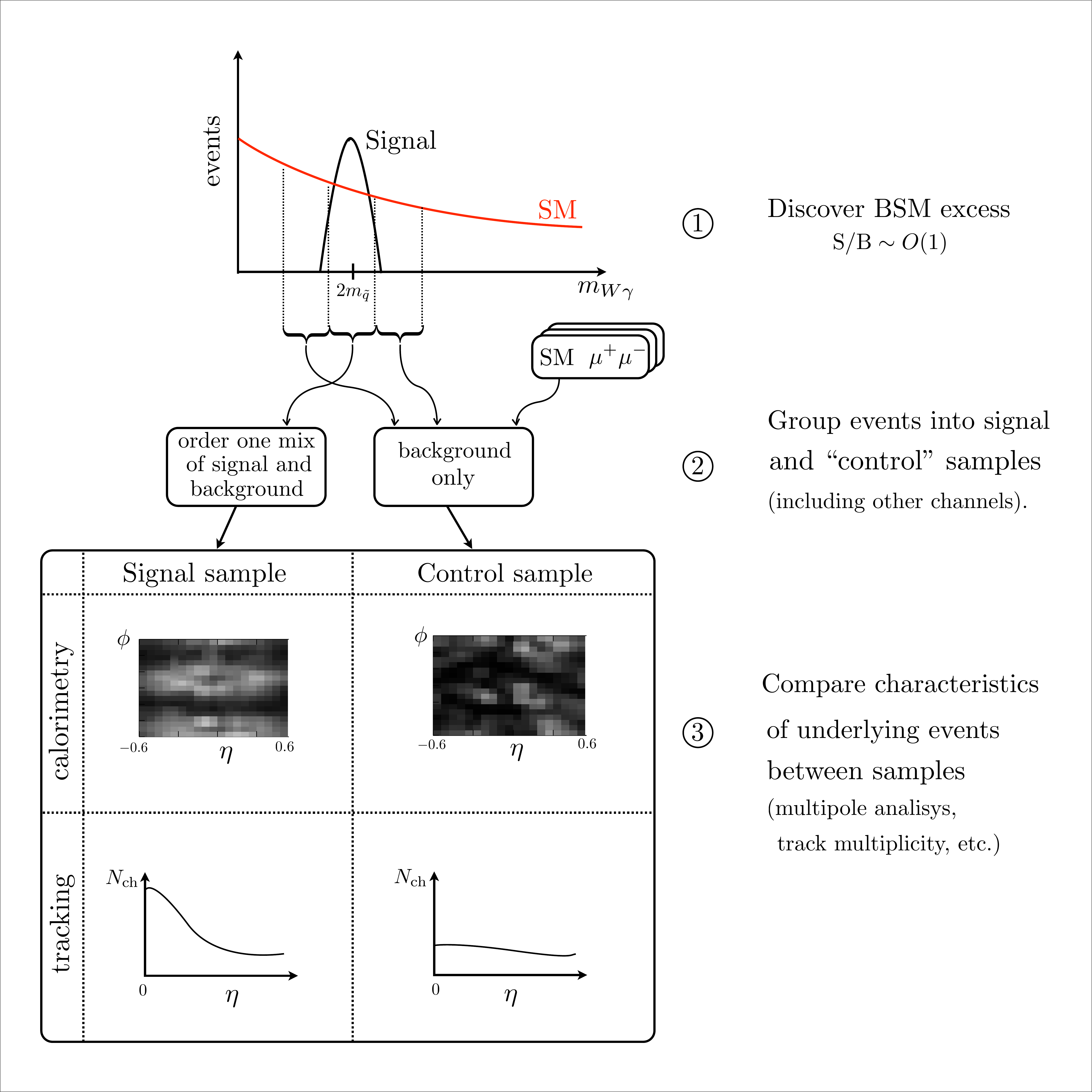}
   \caption{A schematic flow chart of a strategy for searching for anomalous underlying events  
   in our example of squirk production. }
  \label{strategy}
\end{figure}

The goal of the LHC search for this model would be to first establish that new physics is seen using standard hard physics objects emitted in the hard annihilation, and then to extract information about the nature of the new physics. In particular, detection of the unusual ``antenna pattern'' of soft photons in addition to the hard resonance will be a smoking gun signal of the strong dynamics and the presence of a bound state. 
What is a possible strategy to making these discoveries?
In this case the existence of new physics may be demonstrated by a standard hard search. However, the correlation of new physics events with anomalous underlying event may teach us about the nature of the new physics, and perhaps enhance the confidence in the original discovery.     
A rough sketch of a search is shown in Figure~\ref{strategy} and described below.
\begin{enumerate}
\item 
{\bf Establish the existence of an excess of events from new physics}.\\ Signal events will pass triggers with high efficiency due to the hard photon and lepton/jets from the annihilation.
A promising search is to look for a peak in the $W$+photon invariant mass (or rather transverse mass) for leptonic $W$ 
decays~\cite{squirk-ann}. Due to the clean final state and the mass peak a signal-to-background ratio of order 1 may be achieved\footnote{The signal to background ratio of order 1 may be achieved even when the transverse mass peak is smeared due to additional missing energy from hidden glueballs~\cite{squirk-ann}. As will be described later, one may assume glueball emission is suppressed, in which case, the smearing of the peak is reduced and the signal-to-background will improve. In this work we will pick S/B$\sim 1$ but the results can easily be rescaled to other signal-to-background values.}.
\item
{\bf Identification of signal and control samples}.\\
We would like to determine if the new physics discovered in step 1 is associated with an anomalous excess in soft particles. For example, a resonance in $W$+photon can be interpreted as a fundamental heavy $W'$ which contributes no new \emph{soft} physics. If the origin of the BSM excess were a $W'$, the underlying component of the signal events is strictly a long distance phenomena and would thus be similar to the underlying component in SM background events. In order to test whether a new anomalous component in the underlying event appears in correlation with the hard signal, we can compare the underlying part of events in the signal region to the underlying part of events that are known to be background dominated.

One possible background dominated sample of events may be collected from neighboring bins in the kinematic distributions which were used to extract the signal, such as the regions immediately above or below the resonant peak (see Figure~\ref{strategy}). Undelying event studies have shown that the  characteristics of UE's are largely independent of the hard $p_T$ in 
the event~\cite{UE, UE-LHC}. The characteristics of the underlying event can thus be interpolated into the signal region reliably.

In addition, since a large contribution to the underlying event at LHC is not directly related to the primary hard interaction, sample background underlying events may be collected from other channels with similar kinematics. For example, the UE in Drell-Yan $\mu+\mu-$ with a $\sqrt{\hat s}$ that is similar to the discovered resonance, may be studied relatively easily and compared with the signal.
\item
{\bf Comparison of underlying events}.\\
Signals of anomalous underlying events may be discovered either in the tracking or calorimetry of the LHC experiments.
UE studies often use the number density and $p_T$ distributions of soft tracks in the central region (or transverse regions for events with jets), e.g.~\cite{UE,UE-LHC, MB-LHC}.  In the case where the soft new physics is mostly in photons, the distribution of diffuse  unclustered energy in the central EM calorimeter may be more promising. In particular, a multipole decomposition of unclustered energy may be used to recognize antenna-like angular patterns in the underlying event. As we will discuss, even in the case of soft photons a tracking signal may be seen from photons that convert in the tracking system.
In this work we will consider both tracking and calorimetry signals, and present a more  detailed study of the later.
\end{enumerate}

A similar strategy may be applied to other LHC searches, particularly searches for di-lepton, di-jet, or di-photon resonances potentially teaching us that the new resonances are in fact de-excited bound states. An UE analysis of this type may also be applied to searches for large missing transverse energy. In this case an anomalous underlying component which is correlated with missing transverse energy may be a signal of light hidden valley particles that are decaying to soft SM particles and are ``faking'' the missing energy signal. 

In the example above the hard search alone gave a clear new physics signal and the UE study gave complementary information. One can also consider cases where a hard search is not likely to produce a significant excess above SM background (e.g. in di-jet searches). As we will discuss briefly, in that case the anomalous underlying event may be used as a tool to reduce backgrounds and improve the confidence in the initial discovery.

The method of discovering soft new physics will depend on the channel in which it is produced. In the example discussed above, where a large number of soft photons are emitted, discovering the anomalous UE with tracking may be difficult (but perhaps possible) and calorimetry may be more promising. In this work we will focus mostly on this case. 
On the other hand, in some new physics scenarios the new soft physics is hadronic (e.g. in the case of the ``hadronic fireballs'' associated with colored quirks~\cite{quirks}). As we will discuss, in this case both tracking and calorimetry may be useful.

Having sketched the goals of this work we will consider the example above in more detail. In section~\ref{newphysics} we will present a quirk model which may lead to a large number of soft photons in LHC events. We will briefly consider the dynamics and assumptions that lead to such events and parameterize some of the uncertainties. The model is directly motivated by folded supersymmetry, a recently proposed solution to the hierarchy problem using ``superpartners'' that are colored under a new gauge group instead of SM color. 
In section~\ref{detector} we will describe how soft particle, particularly photons, interact in the LHC detectors and how they may be discovered either by tracking or calorimetry. 
In section~4 we focus on the calorimetric signal and present a simulation of a dedicated toy detector as well as simulations of
the signal and background. We propose a possible observable based on a multipole decomposition of unclustered energy in the central calorimeter to discriminate signal and background events.
We finally give two examples of analyses that may be done in searches for anomalous underlying events.
In section~5 we present a brief discussion and conclude.

\section{Quirks - New Physics with a Soft Diffuse Signal}\label{newphysics}

Before presenting a particular model, we will briefly discuss the following question --
what type of soft signals can potentially lead to an observable anomaly in the underlying event? 
Obviously, an observable signal of this type will only exist if some new dynamics is generating a large multiplicity of soft particles which are unclustered. Additionally, an observable signal must dominate over the characteristic UE for the particular hard final state in question. 
Because most of the standard UE comes from QCD, most of the soft energy is emitted in the forward direction. In particular, distributions of energy and track multiplicity are expected to be  flat in pseudorapidity (though this will ultimately be measured). Therefore new dynamics which emits soft particles into the central regions of the LHC detectors are more likely to be observable due to the lower background there. Finally, emission of soft particles in a particular angular pattern can help distinguishing  signal from background. The new physics model we will now discuss will have all of these features.

\subsection{(S)quirks and their Dynamics}\label{quirkydynamics}

We will now present a brief review of quirks, a new type of particle which exhibit new and interesting dynamics. These dynamics have been known for a while~\cite{bj}, and have received recent attention in the context of the LHC as well as their name in~\cite{quirks}. More recent discusions are 
in~\cite{nussinov}. Similar dynamics may also be present in Hidden valley scenarios~\cite{hv1}.

Though quirky dynamics may be exotic, it can arise from a seemingly mundane extension of the SM.  
Consider a model with a new strong force, QCD'. For concreteness we will take QCD' to be an asymptotically free SU(N) gauge force with a QCD' scale $\Lambda$ of order a few GeV. We introduce some matter fields, collectively called $q'$, which are charged under the new force and also under the standard model interactions. As opposed to regular QCD in which there are quarks, $u$ and $d$, whose masses are bellow the QCD scale, we will consider a case where all matter charged under QCD' is more massive than~$\Lambda$
\begin{equation}
m_{q'}\gg\Lambda\,,
\end{equation}
where even a little hierarchy between these two scales will be sufficient.
When this condition is satisfied, the particles that are charged under the new force have been called ``quirks''~\cite{quirks} (or squirks, depending on their spin) for reasons that will become clear presently. 
In particular we will consider two types of quirks -- \emph{uncolored quirks}, which in addition to being charged under QCD' carry only electroweak quantum numbers, and \emph{colored quirks} which also carry SM color. Beyond these two categories one can make various choices, such as the spin of the particles, that will matter only in the details. 
The examples we will consider are motivated by a specific model, folded supersymmetry~\cite{folded}.  
Starting with the uncolored case, consider the scalar squirk $\tilde q'$ with the following quantum numbers
\begin{center}
\begin{tabular}{c|cccc}
\ & SU(N)$_{\mathrm{QCD'}}$ & SU(N)$_{\mathrm{QCD}}$ & SU(2)$_L$ & U(1)$_Y$ \\
\hline
$\tilde q'$ & {\bf N} & {\bf 1} & {\bf 2} & 1/6  \\
\end{tabular}
\end{center}
This squirk doublet is quite  similar to the doublet squarks in supersymmetric extensions of the standard model, with the exception that it is charged under QCD' instead of QCD.   In fact the similarity to squarks may be exploited to solve the hierarchy problem in folded supersymmetric models in which the superpartners of SM quarks are charged under a new QCD' (with N=3) instead of QCD.

Our choice of the QCD' strong scale is also motivated by folded supersymmetric models in which the two QCD scales are related by a $Z_2$ exchange symmetry\footnote{The different spectrum in the two QCD sectors will introduce a small logarithmic difference between the two strong scales through running.}. The lightest hadron in the QCD' sector is the glueball, most probably of the scalar type. This glueball is stable in a pure QCD-like theory, and may decay slowly to SM particles by higher dimensional operators. For our choice of strong scale, glueball decays occur far outside of any particle detector. 

A second particle motivated by folded supersymmetry that we may consider is a colored fermion quirk with quantum numbers
\begin{center}
\begin{tabular}{c|cccc}
\ & SU(N)$_{\mathrm{QCD'}}$ & SU(N)$_{\mathrm{QCD}}$ & SU(2)$_L$ & U(1)$_Y$ \\
\hline
$\tilde g'$ & {\bf N} & ${\bf \bar{N}}$ & {\bf 1} & 0 
\end{tabular}
\end{center}
In folded SUSY such particles are the symmetry partners of the QCD gluon. The dynamics of the colored quirk share some common features with those of the uncolored squirk. In this work we will focus on the later and comment on the colored case occasionally.

In addition to the modification of the hadronic spectrum, our QCD' sector will exhibit qualitatively different dynamics at colliders once squirks are produced~\cite{bj, hv1, quirks,squirk-ann}. 
Squirks will be produced at the LHC via weak interactions by either a Drell-Yan process or gauge boson fusion. They will typically be semi relativistic upon production, with a $\gamma$ factor of order two or so.
The two squirks will fly off back-to-back but once their separation approaches $\Lambda^{-1}$ they will start feeling the effects of QCD' confinement. More specifically the squirks will develop a gluon field configuration which, as the distance between them grows, will turn into a QCD' flux-tube or a QCD' string.
In normal QCD with light matter the QCD string is torn promptly by producing a light quark-antiquark pair which can then separate the string to two (or many) pieces.
Such a hadronization mechanism is possible because the energy density in the QCD string, $\Lambda^2$ is greater than the energy density needed to pair create, of order $m_q^2$.
In quirky QCD such a soft hadronization mechanism is obviously absent (i.e.~exponentially suppressed~\cite{bj}) due to the heavy quirk mass,   $\Lambda^2\ll m^2_{q'}$. 
Furthermore QCD' dynamics will not cause the heavy quirks to loose a significant amount of their kinetic energy, say by showering or fragmentation, over a distance of $\Lambda^{-1}$~\cite{BJ-fragmentation}. 
Instead the two heavy ends of the string will continue to move apart transferring kinetic energy into the QCD' string whose tension is a constant of order $\Lambda^2$.
The squirks will eventually stop when all of their kinetic energy was transfered to the string
\begin{equation}
 \Lambda^2 L_\mathrm{max} \sim E_k\,,
\end{equation}
where $L_\mathrm{max}$ is the maximum string length and $E_k=\sqrt{\hat s}-2m_{q'}\sim m_{q'}$ is the initial kinetic energy in the squirk system upon production. The squirks will then be pulled together by the string, beginning oscillatory motion. We may thus conclude that squirk production at a collider is in fact a production process of a single highly excited mesonic bound state, ``squirkonium''.  As we will discuss in further detail in the next subsection, this excited meson will decay radiatively to the ground state and finally annihilate to hard decay products. For our choice of $\Lambda$ all of the processes above will be prompt on collider time scales.

What is our signal? The soft radiation emitted during the decay from the highly excited to the ground state is the soft signal which we will ultimately detect. In the case of uncolored squirks the signal will consist of many unclustered soft photons\footnote{We will discuss soft hidden glueball radiation in the next subsection.}.  In the colored case the soft signal (which was dubbed a ``hadronic fireball''~\cite{quirks}) will mostly consist of a large multiplicity of pions. The hard annihilation products will provide the hard primary signal.

What are the hard primary signals to search for?
In the colored case: the hard signal was discussed in~\cite{quirks}. The most common visible annihilation is to jets, but given the higher production rate for colored quirks, leptons or photons may be a feasible final state. 

For uncolored squirks: a neutral squirkonium state will typically decay to two hard  (but invisible) QCD' glueballs which will not give a triggerable signal. However, the dominant production of squirks will be via an s-channel $W^\pm$ and thus the produced squirkonium is charged under standard model QED. Its decay products will then always contain a charged particle and leave a visible signal in detectors. An interesting decay channel which may dominate is $W^\pm\gamma$~\cite{squirk-ann}. 
Because the annihilation will typically occur at or near the ground state, a resonance peak is expected in the invariant mass of $W\gamma$.
 
The details of the search for the hard annihilation of uncolored squirks is discussed in~\cite{squirk-ann}.
Here we will ask the following question: Given an excess of $W\gamma$ events with a signal-to-background ratio of order one, can we learn more about the nature of new physics by carefully analyzing the underlying event?  To answer this we will discuss the energy and angular distributions of the soft photon radiation in further detail.

\subsection{Soft Radiation} \label{radiation}

We will now consider the dynamics of a highly excited squirkonium bound state and its decay. Because the bound state is produced with a very high principal quantum number, the system may be treated semi-classically~\cite{quirks}. The system can then be modeled by two heavy particles of masses $m_{1,2}=m_{q'}$  with an attractive linear potential $V=\Lambda^2|\vec x_1 - \vec x_2|$. 
In the center of mass frame the motion is parameterized by the total energy $E$ and by the angular momentum $l$. The energy is initially set by the energy of the collision $\sqrt{\hat s}$ and will decrease as the oscillating system radiates. The angular momentum number $l$ is initially of order one, and will change as the system radiates by one unit per radiated quantum on average. The angular momentum will therefore increase on average by a random walk.  Classically, the angular momentum is related to the impact parameter $b\sim l/m_{q'}$, and is important to determine the likelihood of short distance effects such as re-annihilation as will be explained below. 
 
The classical trajectory of the two particles is easily computed and is to a good approximation linear oscillatory motion. The period of the motion is  
\begin{equation}
T\sim \frac{p_\mathrm{near}}{\dot p} \sim \frac{m_{q'}}{\Lambda^2}
\end{equation}
where $p_\mathrm{near}$ is the the momentum of the quirk at the point of nearest approach (or at production for the first periods) and $\dot p\sim \Lambda^2$ is the force acting on the quirks.

The two particles are charged under standard model QED (with charges $\pm 2/3$ and $\pm 1/3$) and are also triplets under QCD'. The accelerating charged particles will radiate to both of these sectors. As a simplifying first step we will completely ignore radiation of QCD' glueballs, returning to this assumption later. In this simple approximation the excited system will slowly decay by classically emitting electromagnetic radiation in accordance with Larmor's law
\begin{equation}
\dot E = \frac{8\pi \alpha}{3 m_{q'}^2}\,\dot p^2\,.
\end{equation} 
The spectrum and angular distribution of photons may also be readily calculated by Fourier decomposing the retarded potential far away from the source~\cite{jackson}. 
Here we will discuss the qualitative features of the distribution, leaving details for Appendix~A.
The Fourier series will be dominated by the frequency of oscillation of the quirk system which for most of the motion\footnote{As the system looses energy, $p_\mathrm{near}$ will decrease as $\sqrt{E^2-m^2}$ and the frequency will gradually increase, however most of the energy will be radiated at frequencies near the initial one, of order $\Lambda^2/m_{q'}$.} is
\begin{equation}
\omega\sim 
\frac{\pi \Lambda^2}{m_{q'}}
\end{equation}
For a squirk mass of 500 GeV and a QCD' string tension of (5 GeV)$^2$ the photon spectrum is dominated by frequencies of order a few hundred MeV. 
Such a soft photon by itself is not observable at the LHC. however as we will show, if a significant amount of the squirkonium's energy is lost though such photons, they may give an observable modification to the underlying event.

The angular distribution of the emitted radiation is particularly important. A relativistic charged particle which decelerates and is brought to rest is known to radiate in a cone in the forward direction. A particle accelerating from rest radiates in a similar pattern. 
The oscillatory motion may be thought of as a succession of such accelerations an decelerations.   
The ``antenna pattern of radiation'' (see figure~\ref{fig-cartoon}) may be understood as a combination of two such cones back to back. The details of computing this distribution are given in Appendix~A. 
As we shall see later, a calorimeter surrounding the decaying system will thus see two circles of radiation in the $\eta$-$\phi$ plane. This peculiar pattern will be the smoking gun signal of such radiative decays.

Before demonstrating how to search for these patterns we should consider some concerns and caveats. For example in order for a large amount of energy to be emitted in radiation we must ensure that the squirks do not annihilate in a highly excited state. This, however, is unlikely.
To argue this we will adopt the semi-classical formalism of~\cite{quirks}.  
  
The probability for annihilation in a single crossing depends strongly on the angular momentum of the bound system, peaking at low $l$'s. 
Classically it is clear that annihilation will only be likely at low impact parameter, where a naive expectation would be that the annihilation probability will scale like $\sigma_{ann}/b^2\propto l^{-2}$. A more rigorous partial wave analysis gives a much stronger dependence, scaling like  $l^{-2-l}$. 
Given that the angular momentum will grow on average as radiation is emitted, the likelihood of early annihilation is determined by a ``horse-race'' between the annihilation cross section and the radiation rate. Very naively one can expect that radiation would win since the radiation rate is proportional to $\alpha$ and the annihilation cross section scales like $\alpha^2$. However to make a clear determination a more careful estimate is required.

The probability for emitting a photon per period is given by 
\begin{equation}
P_\mathrm{rad}\sim T \frac{ \dot E}{\omega}\sim\alpha\,.
\end{equation}  
and thus a photon will be emitted on average once per $\alpha^{-1}$ crossings. This can be compared with the probability of annihilating at low $l$'s. In our case of charged squirkonium, it was shown~\cite{squirk-ann} that at high relative velocities, right after production, annihilation goes dominantly to SM fermions with a cross section
\begin{equation}
\sigma=N_{\mathrm{QCD'}}N_f v_{\mathrm{rel}}\frac{\pi \alpha^2_W}{48 E^2}
\end{equation}
where $N_f=12$ is the number of SM final states, $\alpha_W$ is the weak coupling constant and $v_{\mathrm{rel}}\sim1$ is the relative velocity. Following~\cite{quirks}, the probability for annihilation per crossing  at high velocities is\footnote{ In \cite{quirks} it was argued that EM radiation is not likely to prevent  a \emph{fermionic} uncolored squirkonium system from annihilating early. This is because in the case studied there $P_{ann}$ was only a factor of a few smaller than $\alpha$.
This, however, highly depends on the details of system at hand. In particular the annihilation cross section for our scalar system is significantly smaller.} 
\begin{equation}
P_{ann} \sim \frac{m_{\tilde q}^2}{2\pi} \sigma v_{\mathrm{rel}} \sim N_{\mathrm{QCD'}}N_f \frac{\alpha_W^2}{284}\left(\frac{m_{\tilde q}}{E}\right)^2 \sim 10^{-4}
\end{equation}
where $2 m_{\tilde q}/E$ is typically of order 1/2.   
Therefore, an order of $10^2$ photons will typically be emitted before the initial probability to annihilate becomes appreciable. However, the change in angular momentum due to the emission of these photons  (by $\Delta l \sim 10$ on average) will significantly reduce the probability to annihilate.

This argument implies the bound squirkonium system will likely decay radiatively to a low lying state before annihilating.  This is fortunate not only because the soft radiation provides an interesting signal, but also because the hard annihilation will occur at a fixed invariant mass, easing its identification independently~\cite{squirk-ann}.

Another caveat we should discuss is the possibility of loosing energy by the emission of QCD' glueballs. In the case we are considering hidden glueballs decay outside the detector and do not lead to an observable signal. In fact, one may argue that glueball emission is likely to dominate over radiation of photons because of the strong coupling in the QCD' sector. However, because of the gap in the QCD' spectrum, glueball emission may be suppressed. In the semi-classical limit the oscilating suqirks would prefer to radiate at  frequencies that are in tune with the oscillation frequency $\omega\sim \Lambda^2/m_{q'}$. No hadron in the QCD spectrum is light enough to be radiated in this frequency and we thus expect photons to dominate over the majority of the oscillation time. When the quirks are close to one another the semi classical limit breaks down and quantum emission of a glueball is possible.
However, the probability for such an emission may be kinematically suppressed as well.
According to lattice studies the glueball mass is expected to be of order~3-4 
times $\Lambda$~\cite{lattice}. Given the small hierarchy between the glueball mass and the QCD scale one may estimate this emission perturbatively in which case the probability for emitting glue is down by $(m_{\mathrm{glue}}/\Lambda)^6\sim 10^{-3}$~\cite{quirks}. We can thus naively estimate that the rate of energy loss due to glueball emission is of order a few times $10^{-3}\Lambda$ per period. 

How does this compare with photons? Recall that the probability for emitting a photon of energy $\Lambda^2/m_{q'}$ per period is of order $\alpha$. Very roughly the competition between photons and glueballs reduces to comparing the kinematic suppression of $10^{-3}$ to $\alpha\Lambda/m_{q'}$. Using $\Lambda=5$ GeV and $m_{q'}=500$ GeV the later is of order $10^{-4}$. We thus estimate that losing 10\% of the energy into soft photons is quite reasonable. This however is a very rough expectation, and various factors may affect the answer in either way. For example, in the large $N_c$ limit glueball emission is known to be suppressed as $N_c^{-2}$, giving an order of magnitude suppression even for $N_c=3$. 
Given the high level of uncertainty in the fraction of excitation energy lost to photons we will take a more phenomenological approach. We will consider two cases: case (1) in which glueball emission is highly suppressed and all of the energy is emitted in photons, and case (2) in which we will only allow 10\% of the energy to be emitted in visible photons and the rest is lost to glueballs. We will asses the prospects of observing a deviation from the standard underlying event in both of these cases, finding interesting results in both.

Finally, we will consider the case of colored quirkonium. The story here is qualitatively similar: quirks are pair produced (with a larger cross section) and form a highly excited bound state. When the ends of the string are colored energy may be lost by radiating light QCD pions. Any kinematic suppression which may be present in the QCD' case will now be absent and we expect pions to dominate the energy and angular momentum loss. In~\cite{quirks} it was argued the produced bound state will ring down to a low lying state before annihilating in a significant fraction of production events. It may thus be feasible to search for the hard annihilation products of colored quirkonium states in association with a large multiplicity of soft pions. These new pions have been dubbed ``hadronic fireballs''~\cite{quirks}, but they may be thought of as a new additional component to the underlying event.
As we will see, the techniques for searching for soft photons will also be useful for the hadronic case.
With this in mind we are now ready to ask how a large number of soft particles, particularly photons, can be detected in an LHC detector.

\section{Soft Photons in an LHC Detector}\label{detector}

In this section we will explain how LHC experiments can detect and measure diffuse soft photons and which parts of the detector are relevant for these measurements. We will find that a large number of soft photons can potentially leave observable signals in both  the tracking system and calorimeters of the LHC experiments. We will then consider the prospects and techniques for observing both
tracking and calorimetric signals and distinguishing them from backgrounds.

\subsection{EM Showers and Soft Signals}\label{shower}

We will now briefly review how soft photons interact with matter triggering EM showers. A useful review of these subjects is in~\cite{pdg}. High energy photons ($\gtrsim$50 MeV) interact with matter dominantly by conversion to an $e^+e^-$ pair. High energy electrons loose energy mostly by emitting bremsstrahlung photons. The successive repetition of these processes is an electromagnetic shower. Because photon conversion and bremsstrahlung emission are related by a crossing symmetry the shower may be characterized by one length scale, a radiation length or $X_0$. The radiation length $X_0$, is the typical distance over which an electron looses an order one fraction of its energy by bremsstrahlung. It is also 7/9 times the mean distance a photon travels before converting. $X_0$ depends on the medium which the particles traverse, and is roughly 9.3 cm in solid silicon. 

As the EM shower progresses, the individual particles in the shower loose energy until a critical energy, $E_c$, is reached. Below $E_c$ other energy loss mechanisms such as ionization dominate, quickly ending the shower. The critical energy is roughly 
$E_c\sim (800\mbox{ MeV})/Z$ in material with an atomic number $Z$. Once a photon or electron have been produced below the critical energy they are promptly absorbed and that fraction of energy is considered ``lost'' or deposited. 

The soft photons we are considering will initiate EM showers in the LHC detectors. For our study we will need to know where inside the detector most of the energy is deposited and what fraction of the energy reaches the calorimeter. 
Given a shower that was initiated by a photon of energy $E_0$ the fractional energy deposition is maximized after a distance
\begin{equation}
x_{max}=X_0\left(\ln\left(\frac{E_0}{E_c}\right)+0.5\right)
\end{equation}

\begin{figure}[t!]
 \centering
  \includegraphics[height=10cm]{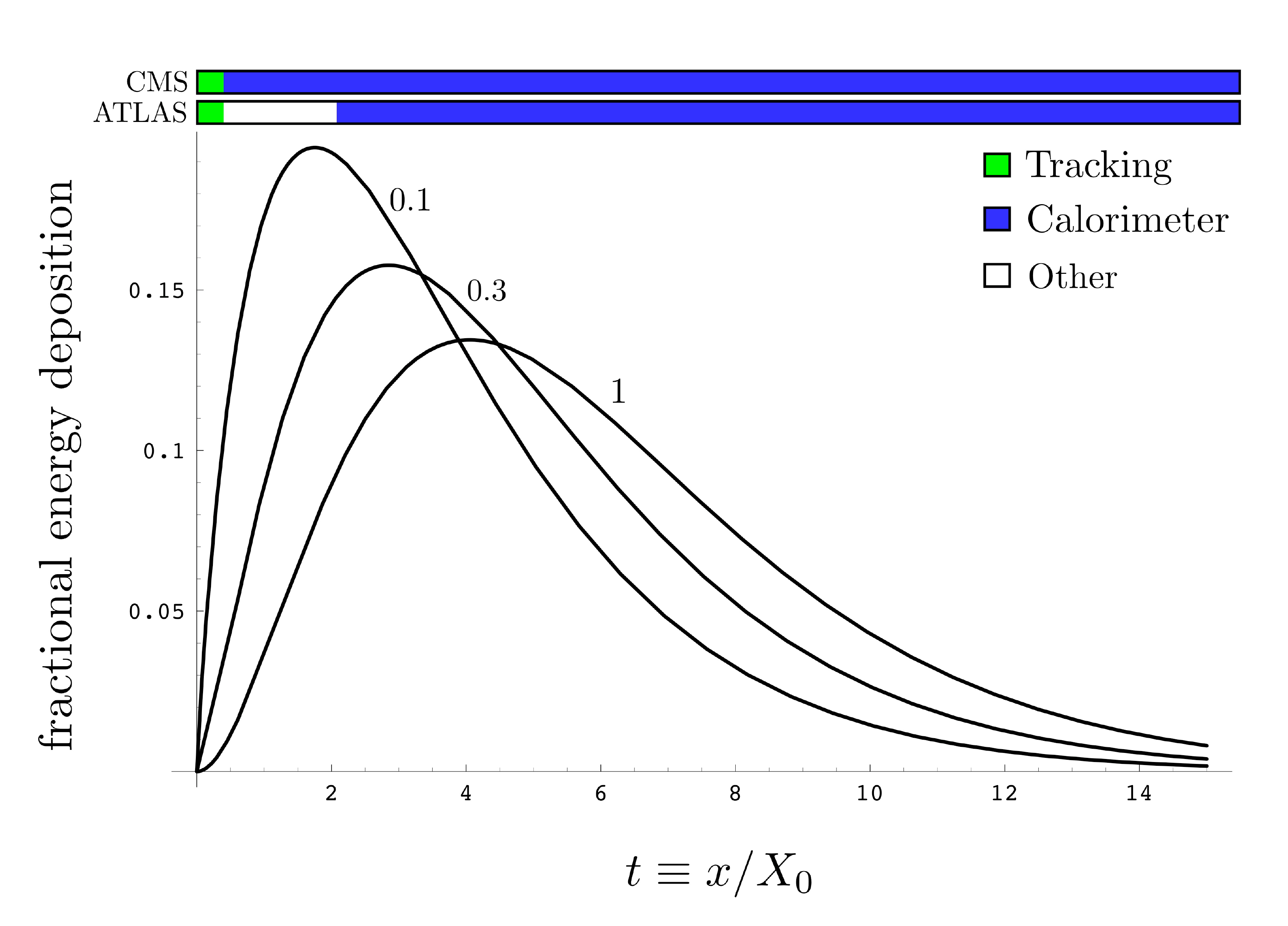}
   \caption{The fractional amount of energy deposited by an electromagnetic shower as a function of distance traveled in radiation lengths for photon initiated shower with energies of 0.1, 0.3 and 1 GeV. A rough estimate of the material budget of the two experiments in the central region ($\eta\lesssim 0.6$)  is shown. 
   }
  \label{figShower}
\end{figure}

The average fractional energy loss of an electromagnetic shower may be expressed as~\cite{pdg}
\begin{equation}\label{eqn:energyLoss}
\frac{1}{E_0}\frac{dE}{dt}=b\frac{(bt)^{a-1}e^{-bt}}{\Gamma(a)}
\end{equation}
where $t=x/X_0$ is the distance in units of radiation lengths. The parameters $a$ and $b$ are defined as $(a-1)/b=x_{max}/X_0$ and $b\sim 0.5$ for our purposes. This fractional energy deposition is plotted in Figure~\ref{figShower} for a shower initiated by a 0.1, 0.3 and 1 GeV photon in silicon.
A rough material budget of the various components of the LHC experiments is also shown.
It is interesting to note a slight difference between ATLAS and CMS. At ALTLAS the calorimeter is behind the coil and the cooling system, which dominate the pre-calorimeter material budget. At CMS the calorimeter is inside the coil. This difference is not crucial for photons in the energy range of interest. We can summarize the matter effects by the following rough estimates
\begin{itemize}
\item Roughly 30\% of the emitted photons undergo their first conversion to an $e^+e^-$ pair inside the tracking system leaving to soft tracks.
\item An order one fraction (30-90\%) of the energy reaches and will be deposited in the electromagnetic calorimeter.
\end{itemize}
This suggests that anomalous underlying events which are dominated by soft photons may be searched for, both in the tracking system and in the calorimeters. In particular, one can search for either or both of the following
\begin{enumerate}
\item A high multiplicity of charge tracks in the central region.
\item Soft energy deposition in the EM calorimeter with distinct geometrical distributions.
\end{enumerate}
Both of these searches should focus on the central regions where backgrounds are low and (for the calorimetric signal) the material budget before the calorimeter is minimal.
We will briefly discuss the tracking signal in the next subsection before going into the calorimetric signal in more detail. 

Before we do so, we will consider anomalous underlying events which are dominated by SM hadrons, as we expect would be generated by colored quirks. In this case one would mostly expect charged and neutral soft pions. The neutral pions, which are roughly one third of the total, will promptly decay to two photons each, producing a large multiplicity of soft photons not unlike the photon signal we have been discussing thus far. The remaining two thirds will consist of soft charged pions which will leave a large number of soft tracks in the tracking systems of the LHC experiments. We thus summarize that both hadronic and ``photonic'' contributions to the underlying event that come from new physics may be discovered by either or both of the following searches: a large multiplicity of soft tracks, and soft and unclustered energy deposited in the calorimeters.

\subsection{The Soft Track Signal}

The efficiency for discovering an anomalous underlying events by tracking depends highly on the tracking system, the algorithm used for tracking and the soft tracking efficiency in the LHC experiments. Here we will merely estimate the expected number of soft tracks from new physics and from backgrounds, and identify potential experimental challenges. 

What is the number of expected soft charged tracks in standard model underlying events? This has been studied extensively for various event generators in preparation for the initial tuning of MC generators during early running of the LHC. The numbers may vary among different event simulations, but it is useful to get ballpark estimate. Eventually these quantities will be measured and compared to signal (as describe in the strategy in section~1).
For example, a study which is useful for our purpose~\cite{UE-LHC} counts the number of soft charged particles in the central region in Drell-Yan di-muon events. For hard di-muon pairs, above a TeV or so, the number of charged particles per unit $\eta$ and $\phi$ with $|\eta|<1$ is about $dN_{\mathrm{ch}}/d\eta d \phi \sim 0.7 - 1.7$, depending on the Monte Carlo generator used. To get the total number of charged particles we multiply by $4\pi$ giving of order 10-20 tracks. At high luminosity one should add a significant number of tracks from pile-up of soft minimum bias events. At $4\times 10^{34}$ cm$^{-2}$ s$^{-1}$ the expected number of min bias events per bunch crossing is of order 10 (a cross section of $\sim100$ mb at 0.1 mb$^{-1}$ per bunch crossing), and more than double that at design luminosity. Each MB event is expected to produce 10-16 charged tracks in the central region~\cite{MB-LHC}. In total, a conservative tally may produce as many as 200 particles per event in the $|\eta|<1$ region.

How does our signal compare? The total number of photons is given by
\begin{equation}
N_\gamma\sim
\frac{\sqrt{\hat s} -2 m_{q'}}{\omega} f_\gamma \sim \frac{2}{\pi} \frac{m_{q'}^2}{\Lambda^2} f_\gamma
\end{equation}
where $f_\gamma$ is the fraction of the energy emitted in photons. We estimated above that roughly 30\% of the outgoing soft photons convert to electron-positron pairs in the tracking system. 
Taking a quirk mass of 500 GeV with $\Lambda=10$ GeV and assuming $f_\gamma$ to be 10\% we get about 200 conversions in the tracking system (and thus about 400 tracks). Due to the geometric pattern it is reasonable to expect an order one fraction of these tracks to be in the central region.
This is a reasonable starting point. The ability to claim that there is an excess of soft tracks from new physics will depend on how well the typical underlying event will be measured, the standard deviation from the average underlying event, and the exact size of the backgrounds, as well as the amount of energy emitted in photons versus hidden glueballs.
In this search one may also search for the geometric pattern of the signal photons. This may be particularly useful since the charged particle density from backgrounds is expected to be flat in the $\eta-\phi$ plane.

We should stress however that the tracking signal that originates in photons may suffer from serious systematic issues. For example, it is not clear that a soft photon that converts somewhere in the middle tracker will be identified as two tracks. The conventional soft track algorithms may reasonably require some hits in the inner-most layers of the tracking system~\cite{UE-LHC}, whereas a displaced photon conversion may ``skip'' these inner layers.
This would reduce the 30\% estimate above to a few percent. This concern perhaps implies that counting tracks may not be the most efficient way to search for our signal. Instead one could hope to construct an observable that that gauges the amount of activity in the tracking system as a function of $\eta$, $\phi$, and perhaps radial depth $r$. The tally of charged tracks above should thus be taken as a rough indication for the feasibility of the search and its challenges. 

For colored quirks, on the other hand the prospects are more promising. Most of the energy is lost by soft pions, an order one fraction of which will be charged (say two thirds) and leave soft tracks. In this case these will be genuine tracks that originate from the interaction point, rather than converted photons. 
In this case the number of pions will be roughly 
\begin{equation}
N_{\pi^\pm}\sim 
\frac{\sqrt{\hat s} -2 m_{q'}}{\omega} \frac{2}{3} \sim \frac{2}{\pi} \frac{2 m_{q'}^2}{3 \Lambda^2}\,.
\end{equation}
For a quirk mass of 500 GeV and $\Lambda\sim 10$ GeV this gives of order a thousand charged tracks. This seams promising, given the expected backgrounds. Here too, however some systematic issues may be important, such as the tracking efficiency at such high multiplicities. Furthermore, investigations with a toy detector simulation (see next section) show that some of the charged pions may reach the EM calorimeters. These soft charged pions, which are highly curved, may ``pollute'' the pattern of soft energy in the Ecal which will be studied in the next section.

\section{Anomalous Underlying Events in the LHC Calorimeters}

In this section we will consider the feesability of observing anomalous underlying events using the distribution of soft energy in the electromagnetic calorimeters. We will simulate background and signal events. Due to the potential sensitivity to matter effects in the detector we simulate a ``toy detector'' using GEANT4 which reproduces the key features required for our analysis. Finally, we will analyze a sample of signal and background events and propose a method to distinguish them using a ``CMB-like'' multipole decomposition. We will propose two possible statistical analyses, one of which is gear towards a signal in soft photons and the other towards hadrons. 

\subsection{ The Signal and the Background} \label{SigBG}

For the purpose of our analysis we generated 500 signal events and 500 background events.\\[8pt] 
\underline{Background:} Our background sample should consist of the underlying part of hard $W$+photon events. A significant portion of this background is expected to come from pile-up of minimum bias which are typical soft QCD events. Therefore, as our background sample we generated 500 \emph{modified} minimum bias events as follows. Min-bias events were generated using PYTHIA \cite{Sjostrand:2006za} tune A, including pile-up at a luminosity of $4\times 10^{33}$ cm$^{-2}$s$^{-1}$ (0.1 mb$^-1$ per bunch crossing). Due to our limited ability to simulate the calorimeter, we modified the minimum bias  by defining \emph{all} outgoing particles to be photons with a momentum that is identical to that of the original particle.
A more in-depth analysis (i.e. one which employs the full detector simulation) would not make this modification. This modification, however, is very conservative in that it makes that background maximally similar to our calorimetric signal. In particular, in unmodified min-bias events soft charged particles will curve in the magnetic field and will not reach the EM calorimeter thus reducing the background in the Ecal.  For this reason, it is important to keep in mind that a more careful analysis may show that the analysis of the upcoming section is underestimating the efficiencies for distinguishing the anomalous underlying event from the SM underlying event.\\[8pt]
\underline{Signal:}
As explained above, our signal is a large number of soft photons with an angular distribution similar to that depicted in Figure~\ref{fig-cartoon}. The details of the angular distribution and frequency spectrum in the center of mass frame are presented in Appendix A. We generated events according to this distribution. The total amount of energy in photons is $(\sqrt{\hat s} -2m_{q'})f_\gamma$ which is typically of order 1-2 time $ f_\gamma m_{q'}$. The fraction $f_\gamma$ accounts for energy lost to invisible glueballs and is taken to be 1 or 0.1 as discussed in Section \ref{radiation} and  $\hat s$ is generated according to the squirk production cross section. The photon distribution is further rotated and boosted longitudinally according to the differential cross section for quirk production. When generating these events we used the MRST parton distribution functions~\cite{MRST}. It is interesting to note that the longitudinal boost did not affect the pattern of photons significantly because the heavy squirkonium is produced near rest.
Obviously, these soft photons will be generated on top of the standard underlying event at the LHC. Therefore, to each signal event we added it own background distribution of soft photons. To do this we added to the signal a modified minimum-bias event which was generated as described above for the background.

\subsection{Simulating a Toy Detector}

The soft nature of our signal makes it particularly sensitive to detector effects.
Several simple detector simulations
exist and are freely available -- examples of these are ATLFAST \cite{atlfast} and PGS \cite{pgs}.  However, none of these correctly capture the low energy physics 
relevant to the processes we are considering.
The crucial question is: given a particular distribution of photons, how much of the energy reaches the calorimeters and to what extent is the angular pattern recognizable? 

The back-of-the-envelope estimate of section~\ref{shower} shows that indeed a significant amount of energy will reach the calorimeters. Understanding the effects of showering in matter and a magnetic field on the pattern, as well as a systematic comparison with backgrounds require simulating the events and the detector. To do this reliably one would need to know how much material each photon traverses on its way to the calorimeter and how it showers. This, of course, is what the full detector simulations do.
However, for the purpose of our initial theoretical study we can ignore the details of the inner detector. Instead we chose to roughly reproduce the material budget of a ``typical'' detector at various distances. 
For this we constructed a hypothetical toy detector  (we arbitrarily designed it  ATLAS-like) in GEANT4 \cite{Agostinelli:2002hh}, a general package for modeling the passage of particles through matter, putting together a toy tracking system and calorimeter.

A cross section of our toy detector is shown in Figure 4.
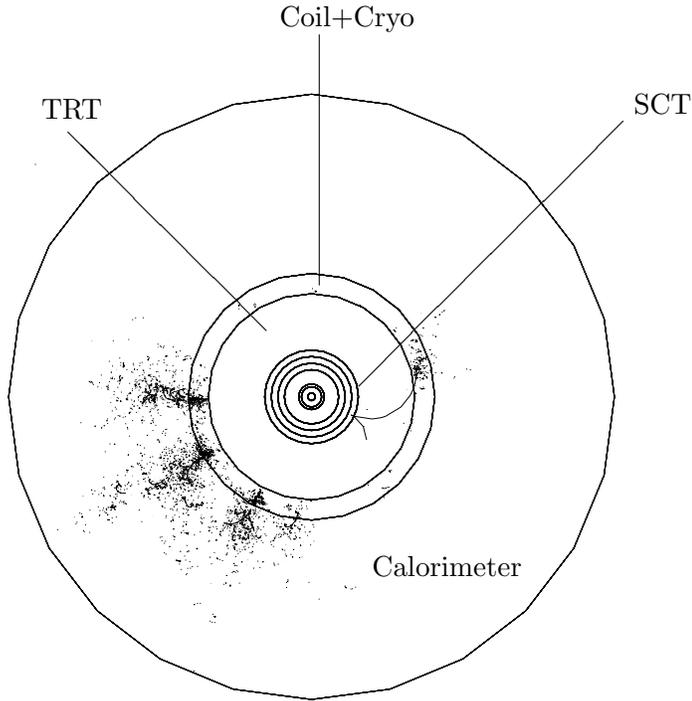
\begin{figure}[t!]
  \centering
   \includegraphics[height=12.5cm]{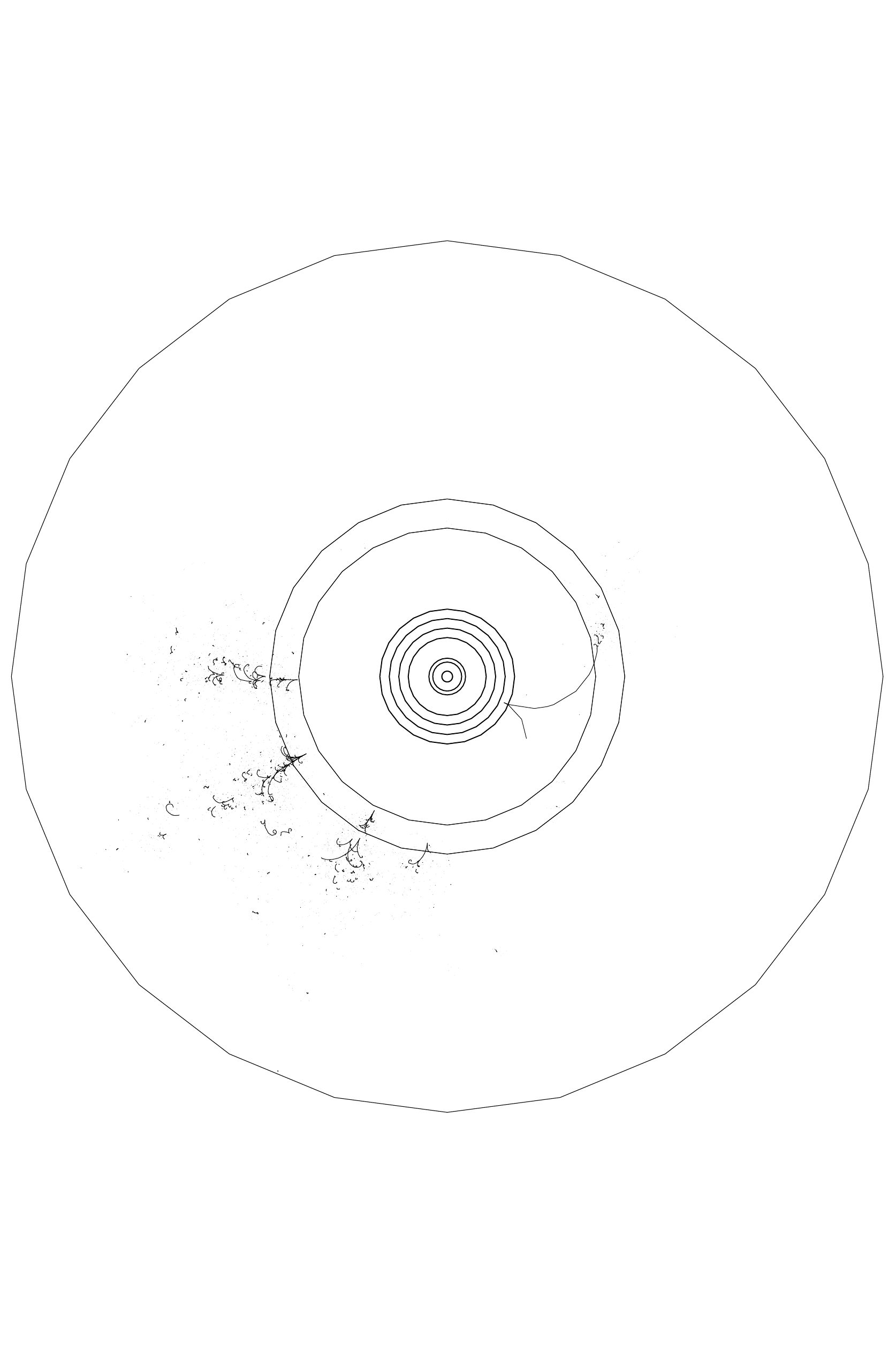}
  \put(4,285){SCT}
  \put(-100,182) {\line (1,1) {100}}
  \put(-220,283){TRT}
  \put(-135,203) {\line (-1,1) {75}}
  \put(-130,318){Coil+Cryo}
  \put(-115,220) {\line (0,1) {95}}
  \put(-95,110){Calorimeter}
  \caption{A cross section of the toy detector used for our analysis.  This simple detector (inspired by ATLAS) is sufficient to answer the simple questions -- how much electromagnetic energy is expected to reach the various parts of the calorimeter?  and is the angular pattern of soft energy visible? Five sample showers are displayed above, originating from five low energy photons.  Evidently, four of these photons showered in the coil and one converted to $e^+e^-$ in the tracking system. Only charged particles are shown.}
  \label{figDetector}
\end{figure}
We model only the barrel region of the detector, extending out to 
$\eta\approx0.6$. We focus on the barrel region because the material budget before the calorimeter is significantly increased for the end caps. 
The inner tracker of our toy detector is represented by concentric cylinders of silicon -- these represent the pixels and SCT of ATLAS, but for us
the division into
pixels and micro strips is ignored. Beyond the inner tracker is an outer tracker consisting of Argonne gas representing the ATLAS TRT. The total size and material budget of our tracking system is similar to that of ATLAS.  Between the TRT and the calorimeter lie the coil
and cryostat.  This region is responsible for most of the 
energy loss between the interaction vertex and the calorimeter. 
Ignoring the detailed structure of these components, we modeled them by
a single cylinder of aluminum with a radial width normalized to
produce the correct material budget in front of the calorimeter.
The calorimeter itself is modeled by a single layer of liquid argon cells
divided along the $\phi$ and $z$ axes. Finally we placed the entire 
system in a 2T magnetic field.
Our toy calorimeter is segmented into 20 bins in the z direction, ranging from -800cm to 800cm and 20 bins in the $\phi$ direction. Our binning is quite arbitrary.  The actual analysis of the diffuse energy deposition in the ATLAS calorimeter will involve the identification of energy clusters using a Topocluster algorithm\footnote{This algorithm for collecting data from the calorimeters is different than the usual ``sliding window'' algorithm for identifying photons. In particular the topocluster has a much lower threshold, of order 100 MeV in the barrel region, which is set by noise~\cite{sven}.}  However, such an analysis is beyond the scope of this paper.  

As an example, in Figure 4 we also show the propagation of five low energy photons through
the inner detector to the calorimeter.  The multiple scattering of the particles,
as well as the bending of charged particle trajectories are evident.  Note that only one 
of the photons showered in the tracking system while the majority showered in 
the coil and cryogenic system.  This is consistent with the relatively low material budget in the tracking system.

While this toy detector does not, by any means, replicate the full architecture
of the ATLAS detector, it does capture the most important points.  First,  
the energy loss through pair production of primary photons and subsequent
radiation and absorption of the secondary particles is correctly reproduced.
We verified this with a uniform distribution of photons. In that simulation, the average
energy loss per photon was in good agreement with Equation \ref{eqn:energyLoss} as well
as with an equivalent run of the full ATLAS detector simulation\footnote{We thank Elliott Cheu for this important cross check.}.  Second, the EM showers produced by the photons as they propagate through the detector are fully simulated by GEANT4, producing a realistic angular 
smearing of the energy deposition. 

\subsection{Results}\label{analysis}

\begin{figure}[t!]
  \centering
  \includegraphics[height=12.5cm]{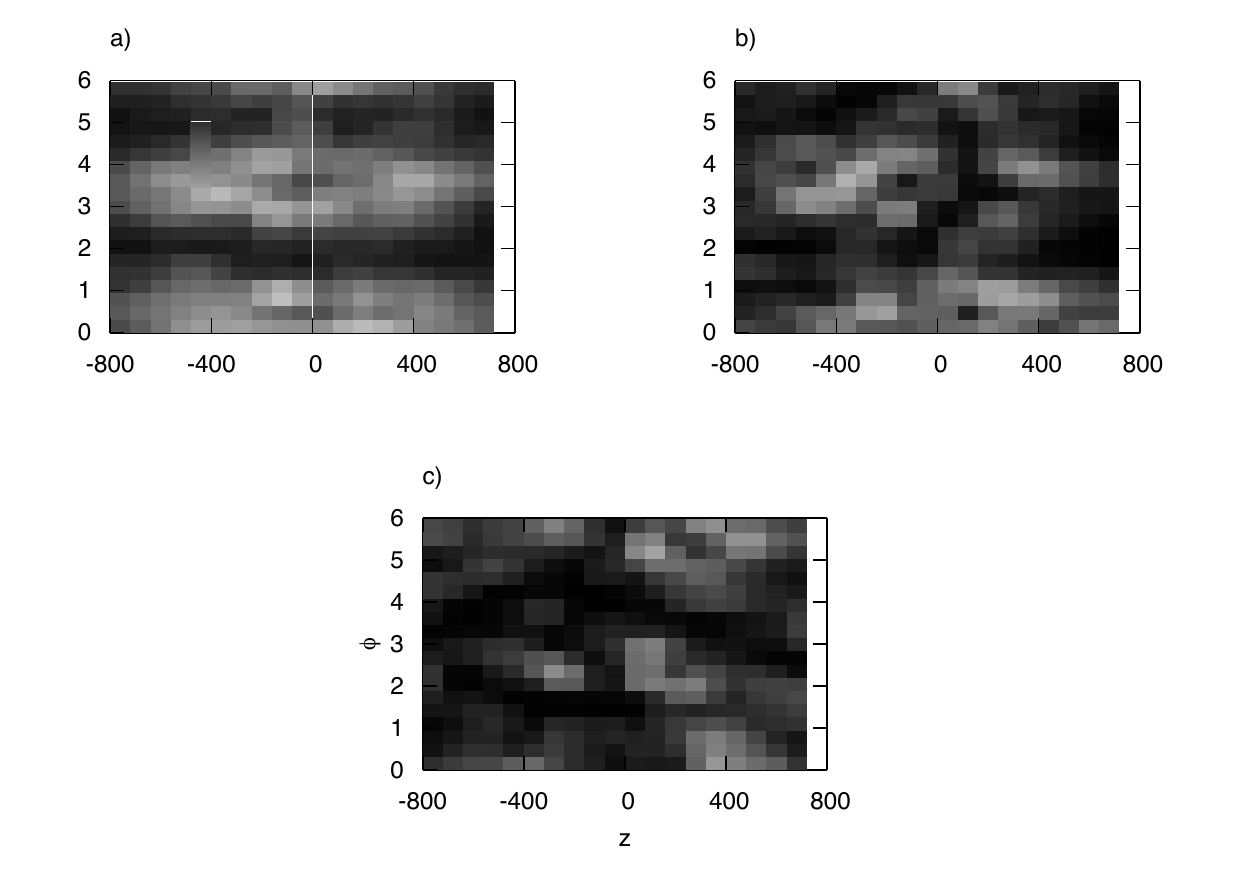}
  \caption{Calorimeter energy deposition in the toy detector simulation. The distribution
  is shown for (a) bound state radiation with 100\% of the energy released in photons, (b) bound state radiation with 10\% of the energy in photons and (c) a  minimum bias event. Brighter squares indicate a higher energy deposition in the cell, however, the scale itself is arbitrary for each figure separately.}
  \label{figCaloEdep}
\end{figure}

In Figure \ref{figCaloEdep} we show energy deposition in the calorimeter for three sample events.  The first is a signal event for which all the bound state energy was emitted in photons. The second is a similar event but with only 10\% of the energy in photons, and the third is a (modified) minimum bias event. The generation of these events is described in section~\ref{SigBG}.   
The antenna pattern, two back-to-back ``doughnuts'' is clearly visible in Figure~\ref{figCaloEdep}(a), where all of the energy is emitted in photons. When only a tenth of the energy goes to photons the pattern is degraded to two back to back blobs and a quantitative statistical analysis, as described below, may be needed to label such an underlying event as an anomalous one. 

One could have worried that any angular pattern in the calorimeter would be smeared by magnetic field effects. However we already see that this will not be the case. This is because the soft photons may be divided into two rough groups -- ones that convert in the tracking system and those that do not. The $e^+e^-$ pairs that are produced by the early converting photons are highly curved by the magnetic field and thus barely deposit any energy in the calorimeters\footnote{This is especially true at ATLAS where the coil is before the calorimeters.}. On the other hand, the photons that contribute most to the calorimetric signal are the ones that pass the tracking system without converting and are unaffected by the magnetic field. 

In order to claim an excess of anomalous underlying events within the signal region one would like to quantify the most atypical features of the signal events, perform cuts on the data and show a statistical excess compared to the expected background. One simple possibility is to cut on the total amount of unclustered energy in the barrel region, $\eta<0.6$. When all of the energy is emitted in photons, the average amount of energy deposited in our toy calorimeter is approximately 550 GeV for a squirk mass of 500 GeV. For comparison, in the average (modified) min-bias event the average was below a 100 GeV. Our modification of the min bias events (see section~\ref{SigBG}), which was geared toward generating conservative backgrounds for pattern recognition (see below), may have increased the later number, but it may be taken as a ballpark figure. 

\begin{figure}[t!]
 \centering
    \includegraphics[height=8cm]{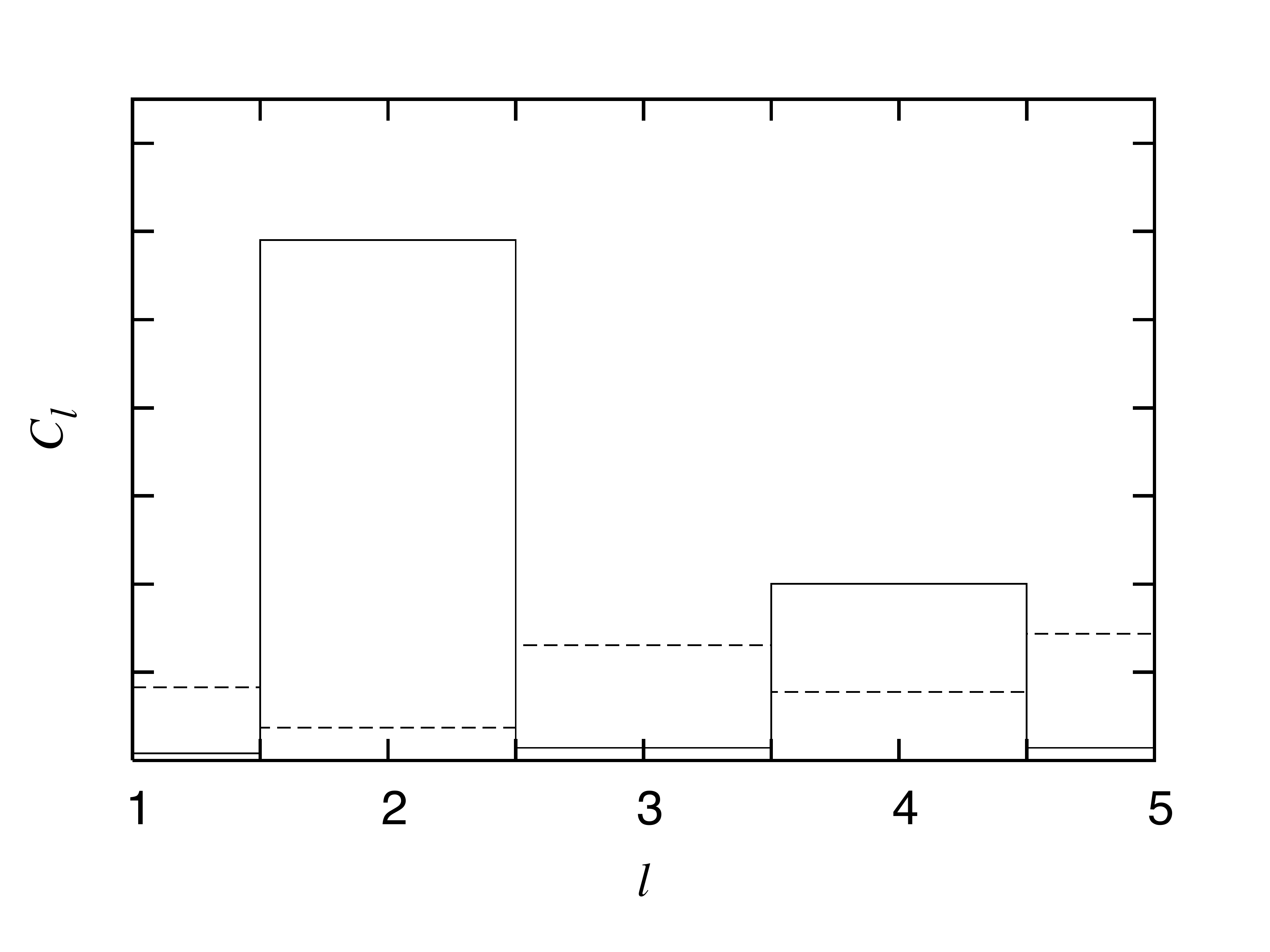}
   \caption{Multipole expansion of calorimeter energy distributions. The multipole 
    coefficients are shown for
    the bound state radiation (solid) and the minimum bias event (dashed) from Figure \ref{figCaloEdep}. The normalization of the y axis is arbitrary.}
  \label{figDecomp}
\end{figure}

In this work we will focus on amore distinct ``smoking gun'' feature of our signal, the angular ``anntena'' pattern of soft energy. Identifying this pattern provides a unique data analysis challenge since most triggering and clustering algorithms are geared toward the identification of hard objects.  
A promising way to quantify the angular correlations of any function defined on a 2-sphere is to use a multipole decomposition, as was shown to be very effective in studies of the cosmic background radiation\footnote{This analogy may be taken further, comparing the subtraction of the galactic plane to the removal of the end caps in our analysis.}.   
Given
a function $f(\theta,\phi)$ The $l^{th}$ multipole coefficient is
$$
C_{l} =\frac{1}{2l+1}\sum_m\left|{a_{lm}}\right|^2
$$
where
$$
a_{lm}=\int d\Omega f(\theta,\phi)Y^{m}_{l}(\theta,\phi)^*.
$$
The coefficient $C_l$ receives its main contribution from fluctuations
on angular scales of ~$\pi/l$.  Thus for the radiation pattern of Figure
\ref{figCaloEdep}(a) we expect the $l=2$ coefficient to dominate.  This is exactly the effect
that is visible in Figure \ref{figDecomp}, with higher multipole moments providing 
sub-dominant contributions to the expansion.  The dashed line in Figure \ref{figDecomp} represents the multipole decomposition for the minimum bias event shown in figure \ref{figCaloEdep}(c).  The difference is quite evident.  

While the difference in the multipole decomposition between specific signal and minimum bias events is striking, it is not statistically significant on it's own.  In fact, the minimum bias events exhibit a wide variation and many of them are atypical and could be mistaken for bound state radiation. In particular, it is known (and confirmed by our background event sample) that some underlying events will contain two broad back-to-back jets  which may also create a peak at $l=2$, as well as higher even $l$'s.  It is thus useful to quantify the expected peak at $l=2$ and look for an excess of signal above background.

There are many observables which can be constructed from the first few multipoles. One possibility which we use for our statistical analysis is the following variable:
\begin{equation}
\rm{sin}{\alpha} =\frac{\sqrt{\left<C_l^2\right>-\left<C_l\right>^2}}{\sqrt{\left<C_l^2\right>}},
\end{equation}
where the averages are taken over the first five multipoles. If we consider these multipoles to be the components of a vector in a five-dimensional space, then $\rm{sin}\alpha$ is precisely the sine of the angle between this vector and the diagonal passing through the origin. An event in which all multipoles have the same magnitude will have $\sin\alpha=0$ whereas an event which is purely in a single multipole has $\sin\alpha=1$.  Hence, this variable measures the degree to which the multipoles tend to differ from each other.  Since the signal events are characterized by large differences between the $l=2$ and the other multipoles, we expect $\rm{sin}\alpha$ to be larger compared to the minimum bias events which tend to be flatter. 

\begin{figure}[b!]
 \centering
     \includegraphics[height=9cm]{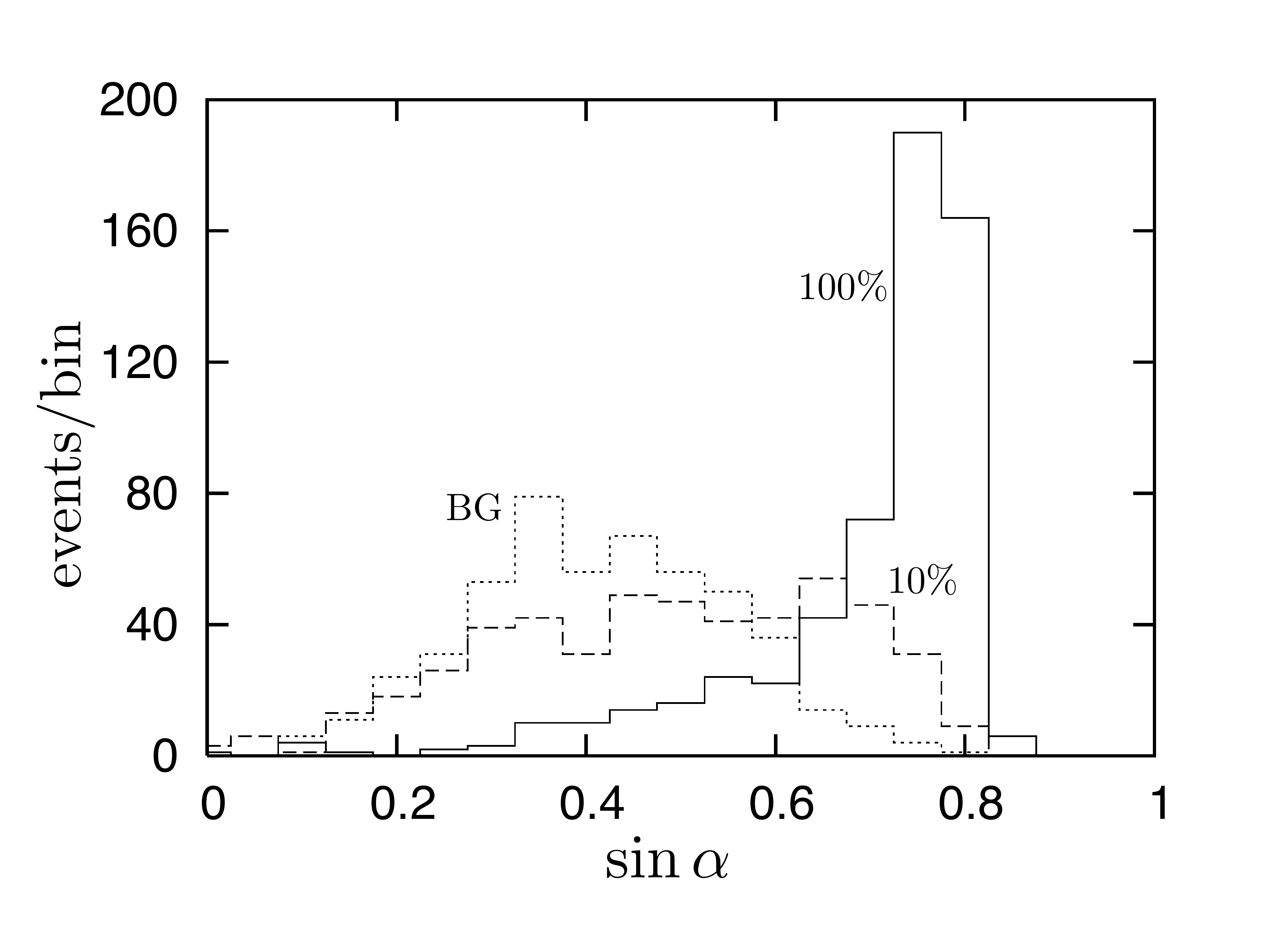}
   \caption{sin$\alpha$ for samples of 500 events: signal events with \%100 of the energy emitted in photons (solid), 10\% of the energy in photons (dashed) and minimum bias events (dot-dashed).}
  \label{figSinAlpha}
\end{figure}

In Figure \ref{figSinAlpha} we compare the $\rm{sin}\alpha$ values for three samples of 500 events. 
The first sample consists of signal events (overlayed on top of background UE's), with all the energy of the bound state emitted in photons (solid), the second consists of similar events for which only 10\% of the energy was emitted in photons (dashed) and the third are the background events (dotted).  
Imposing a cut of $\rm{sin}\alpha>0.7$ we find efficiencies for the three event samples as listed in Table~1.
\begin{table}
\begin{center}
\begin{tabular}{|c|c|}
\hline
\% of energy & Efficiency for\\  
in photons &  $\sin\alpha>0.7$ \\
\hline
100 & 0.74\\
10 & 0.17\\
Background (0) & 0.02 \\  \hline
\end{tabular}
\end{center}
\label{tab-eff}
\caption{Efficienciey for passing a cut of $\sin\alpha>0.7$ for the three event samples of 100\% of the energy in photons, 10\% of the energy in photons, and standard model backgroud. The variable $\sin\alpha$ measures to what extewnt the angulard distribution of soft energy in the calorimeter is dominated by a single multipole.}
\end{table}
The variable we used above does not single out any particular multi-pole. Other possible observables that focus on $l=2$ might be used, though we did not find that these did much better in our particular case.

\subsection{Analyses I: Discovering Anomalous Underlying Events}
Given the efficiencies above, one may analyze the data by several approaches depending on the efficiency and confidence level of the original hard search and on the amount of energy emitted in soft radiation with each new physics event. We will begin with a scenario that is quite likely in our case of uncolored squirks annihilating to $W\gamma$. This hard final state is quite distinct and the signal to background ratio may well be of order one, as demonstrated in~\cite{squirk-ann}. However, even when new physics is discovered we would like to extract as much information about it as possible. In other words, would like to test whether the discovered signal is a ``normal'' resonance in $W\gamma$ or perhaps a quirky bound state. We would like to test whether the new signal is correlated with anomalous underlying events.

To test this hypothesis we may exploit the fact that the underlying event is believed to be quite universal, particularly for similar final states. Following the strategy outlined in section~\ref{intro} we can thus collect a ``control'' data set of standard underlying events by considering similar final states in different kinematic regions, or completely different final states (e.g. di-muon events). 
In fact, the characteristic properties of underlying events, such as charged particle density and $p_T$ distributions have been shown to be independent of   $\sqrt{\hat s}$ in jet events as well as Drell-Yan (beyond a certain energy)~\cite{UE-LHC}. This implies that one can collect a sizable pool of standard underlying events which are not ``contaminated'' by contributions from new physics such as quirks. This large control sample can then be compared with the underlying events in the signal region.

Due to the large number of events, such an analysis may be effective in discovering quirky dynamics even under the more pessimistic assumption of 10\% of the excitation energy being released in photons.
Given enough statistics the the difference between the 10\% and BG distributions of figure~\ref{figSinAlpha} may be established.

To see this lets do a gedanken-analysis with 100 fb$^{-1}$ of data and a simple two-bin chi-squared test. 
In line with the analysis of~\cite{squirk-ann} we will assume the quirk production cross section is of order 10 fb and the branching ratio to leptons times efficiency for the search is of order 0.1,
giving roughly 100 signal events. These numbers are consistent with a squirk mass of 500 GeV, which was used in our simulations. With these efficiencies the signal to background ratio is of order 1, providing clear evidence for new physics well beyond 5 sigma. Thus the ``signal sample'' consists of 200 events, a half of which are true signal. 
The multipole variable $\sin\alpha$ in signal and background events follows the distributions shown in figure~\ref{figSinAlpha} (of course, a more detailed analysis of the UE background and the detector may yield different results). 
As a simple example consider just two bins of  $\sin\alpha$ bellow and above 0.7. Our simulation showed that the fractions of events with $\sin\alpha>0.7$ are 0.02 and 0.17 for background and signal events (10\% of energy in photons) respectively. On average, about 20 of the 200 events will fall in the $\sin\alpha>0.7$ bin. 

This may be compared with a large control sample consisting of pure background underlying events. 
In a control sample of 3000 events only 60 events will fall in the $\sin\alpha>0.7$ bin on average. 
If the measured frequencies landed close to these averages a chi-squared test of independence  gives a very ``bad fit'' of $\chi^2\gtrsim 40$ for a single degree of freedom. Such a measurement would support that the two samples are drawn from two different distributions with a high confidence level.

The ability to consider larger control samples is important to stabilize such a result against statistical fluctuations. For example, consider the case in which we are unlucky and the actual measured numbers of events in the $\sin\alpha>0.7$ bin fluctuate by 2 sigma down and up for both the signal and the control samples respectively. Even in this unfavorable case the measured signal and control samples are statistically independent with $\chi^2\gtrsim 5$ for a single degree of freedom. This demonstrates that in order to carry out this program and large control sample of standard model underlying events should be collected and studied.

\subsection{Analysis II: Improving the Confidence of Discovery of New Physics}

We now consider a different situation (which is not necessarily applicable in our case of uncolored squirks going to $W\gamma$). Lets assume that an order one fraction of the excitation energy (of order 0.5 TeV) is released in photons. However, we can also suppose that the confidence level of discovery in the standard hard analysis is poor, bellow 3 sigma. In this case we can exploit the distinct soft signal (as seen in the 100\% curve of figure~\ref{figSinAlpha}) to improve the confidence level of discovering new physics. Instead of searching for the hard final state $X$, we should perhaps search for $X$+an anomalous underlying event. Such an analysis may fit well with the case of colored quirks in which the amount of soft energy is perhaps larger (since the competition with invisible glueballs is less of an issue) and the hard final state may be in jets, which suffers from a larger background than the distinct $W\gamma$. As we pointed out earlier the systematics of soft hadrons may be somewhat different than those of and should be considered separately.

To demonstrate this lets again assume the efficiencies derived from figure~\ref{figSinAlpha}. We can now use the atypical shape of the underlying event as a discriminant between signal and background. In Imposing a cut of $\sin\alpha>0.7$ the background will be efficiently reduced to 2\% of that in the original search whereas the signal is reduced by $\sim 0.74$. Therefore, in the more optimistic scenario of 100\% of the energy emitted in photons, the original S/$\sqrt{\mathrm{B}}$ may be improved by a factor of  $0.74/\sqrt{0.02}\sim 5.2$. In this case, our underlying event analysis may potentially take a 1-2 sigma bump into a discovery beyond 5~sigma! The less optimistic case of  10\% of the energy in photons the improvement factor is a more modest 1.2.

\section{Discussion}

In this work we have focused on a specific example of new physics, uncolored squirks, which lead to a diffuse soft signal at the LHC. As we mentioned throughout the paper colored quirks may also be promising candidates for generating soft signals in the spirit discussed here. Are there any other cases?
The Hidden Valley scenario~\cite{hv1} has been proposed recently as an interesting possibility for novel LHC signals. The general idea involves the production of new particles which are charged under a new strong force. These particles will shower and hadronize according to the strong dynamics of the new sector. Such a showering process will typically distribute the high energy of the produced particles into many soft valley hadrons, some of which will decay to soft SM jets, leptons or photons. 
One promising way to produce a diffuse soft signal in the spirit of the ones we are studying is to consider an unparticle scenario~\cite{unparticles} in which the conformal symmetry in the hidden sector is broken at some low scale\footnote{This scenario may thus be called a hidden valley~\cite{matt}.}. The strong conformal dynamics will produce very broad ``jets'' of valley particles (the part of the spectrum just above the gap, which is likely to contain resonances) and thus broad distributions of soft and diffuse SM decay products. Indeed, ``multi-unparticle '' production may be enhanced in these scenarios~\cite{feng}.

One particularly interesting possibility is that the hidden sector couples to the SM through the ``Higgs portal''~\cite{higgsportal,hv2}. For example, the Lagrangian may contain an operator 
\begin{equation}
\mathcal{L}\supset |H|^2\mathcal{O}_V
\end{equation}
where $\mathcal{O}_V$ is an operator in the new strong sector. If the hidden sector indeed contains strong and nearly conformal dynamics, the most likely decay channel of the Higgs may be to broad and soft ``jets'' of valley hadrons. The decay of the light valley particles to standard model states will produce a soft modification to underlying events at the LHC. A possible search would be for a single boosted Z plus an anomalous underlying event from the Higgs decay. The multiploe expansion of soft energy in this case would presumably exhibit a peak at $l=1$ with subleading peaks at higher odd $l$'s.

Another interesting possibility which has been proposed  recently is that the Higgs itself is part of the unparticle sector~\cite{TerningUnHiggs}. 
In this scenario the Higgs is part of a strongly coupled sector which also contains very light degrees of freedom. These light degrees of freedom may couple back to the standard model, e.g. by integrating out the heavy Higgs. It is thus plausible that as a Higgs is produced a shower of light degrees of freedom is emitted which consequently decay back to soft SM particles. Such an event will contain the hard decay products of a Higgs in addition to a modified underlying event. In these scenarios, the strategies outlined in this work may ultimately shed new light on the nature of the Higgs.

In conclusion, in this work we have proposed a new signal of new physics at the LHC. Searches for new and heavy physics has rightfully focused on hard physics object. We have demonstrated that new physics may also be discovered in the soft, or underlying component of LHC events. Discovery of an anomalous underlying event may be a unique way to discover the non-perturbative nature of new physics that is otherwise unattainable. We demonstrated our strategy in a simple model of quirks (or squirks) which is inspired by folded supersymmetry, a model for addressing the electroweak hierarchy problem. We have discussed some of the strategies to search for anomalous underlying events, both in the tracking and the calorimetry of the LHC experiments.

A general lesson to be drawn from this work is that underlying event studies should accompany some of the standard searches of new physics.  Such an underlying event study may either enhance the confidence of discovering new physics, or teach us new information about new physics. The hard searches which may benefit from an accompanying UE analysis include searches for heavy resonances, (di-boson, di-lepton or di-jet) as well as searches involving missing transverse energy.


{\bf Acknowledgments:} We would like to thank
Claudio Campagnieri,  
Zackaria Chacko,
Tami Harnik,
 Markus Luty, 
 Shmuel Nussinov,
 Michael Peskin,
 Matt Strassler and 
 Jay Wacker
for valuable discussions.
We would like to especially thank Elliott Cheu of the ATLAS group in Arizona for helpful communication and sharing early results of his simulations. RH would like to thank the Aspen center for physics for hopspitality during various stages of this work. This work was supported in part by DOE grant DE-AC02-76SF00515.

\appendix

\section{Classical trajectory and Photon Spectrum}

To generate the photon spectrum in our simulation we modeled the radiative decay of squirkonium by the classical radiation of accelerating charges. The charged squirkonium is modeled as two particles of equal masses and of charges $\pm2/3$ and $\pm1/3$ with a constant force acting between them.  The classical equation of motion is
\begin{equation}
\frac{dp}{dt}=\Lambda^2
\end{equation}
in the center of mass frame.
Since the squirks are produced semi relativistically we use relativistic momentum $p=m\gamma\beta$.
Assuming the motion is along the $x$ direction the solution is a periodic trajectory with a period of 
\begin{equation}
T=\frac{\sqrt{E^2-4m^2}}{\Lambda^2}
\end{equation} 
with the trajectory for $0\le t<T/2$ described by
\begin{equation}
\label{orbit}
x(t)=\frac{E-\sqrt{E^2 - 2 \Lambda^2 \sqrt{E^2-4m^2} t +4 \Lambda^4 t^2}}{2\Lambda^2}
\end{equation}
where $E$ is the total energy of the system, initially set to $\sqrt{\hat s}$.
The classical photon spectrum may be evaluated by Fourier decomposing the retarded field far away from the oscillating charges. Because of the periodic nature of the motion the field decomposes into a discrete Fourier series with frequencies $\omega_n=n \omega_0$ and $\omega_0=2\pi/T$. Following~\cite{jackson} , the power radiated per solid angle in the $n$th frequency mode is
\begin{equation}
\label{spectrum}
\left(\frac{dP}{d \Omega} \right)_n =\frac{\alpha \omega_0}{\pi}\left|
\sum_i q _i \int_0^{2\pi/\omega_0} dt (
\vec {\rm n}
\times
(\vec {\rm n} \times\vec\beta_i)) e^{in\omega_0(t+R_i(t))}
\right|^2
\end{equation}
where the $i$ sum is over the two charges, $q_i$ is the charge of the particle $i$ and $R_i(t)$ is the instantaneous distance of a far away point from that particle. Assuming the linear motion of equation (\ref{orbit}), the Fourier integral over the path of the particle may easily be done numerically.

To simulate the radiative decay of a quirkonium system produced with a center of mass energy of $\sqrt{\hat s}$ we simply generate a photon according to the spectrum of equation~(\ref{spectrum}) using a trajectory of that energy. We then subtract the photon's energy from the oscillating system and generate the next photon with the reduced energy, and so on. 
As discussed in Section~2, one should assume some of the energy is radiated in the form of (hidden) glueballs. For example, as a benchmark we also consider the case where only 10\% of the energy is radiated in photons. To generate these events we follow the same procedure, but only considering every tenth photon as visible, throwing 90\% of the photons away.


\begin{thebibliography}{99}


\bibitem{hv1}
  M.~J.~Strassler and K.~M.~Zurek,
  Phys.\ Lett.\  B {\bf 651}, 374 (2007)
  [arXiv:hep-ph/0604261].

\bibitem{hv2}
  M.~J.~Strassler and K.~M.~Zurek,
  arXiv:hep-ph/0605193.

\bibitem{hv3}
  M.~J.~Strassler,
  arXiv:hep-ph/0607160.

\bibitem{stopping}
  A.~Arvanitaki, S.~Dimopoulos, A.~Pierce, S.~Rajendran and J.~G.~Wacker,
  Phys.\ Rev.\  D {\bf 76}, 055007 (2007)
  [arXiv:hep-ph/0506242].
  
\bibitem{split}
  N.~Arkani-Hamed and S.~Dimopoulos,
  JHEP {\bf 0506}, 073 (2005)
  [arXiv:hep-th/0405159].
  
  
\bibitem{UE}
  R.~D.~Field  [CDF Collaboration],

\bibitem{UE-LHC}
D.~Acosta, F.~Ambroglini, P.~Bartalini, A.~De Roeck, L.~Fano, R.~Field and K.~Kotov,
CERN-CMS-NOTE-2006-067;
%

\bibitem{MB-LHC}
A.~Moraes, C.~Buttar and I.~Dawson,
  Eur.\ Phys.\ J.\  C {\bf 50}, 435 (2007).
  


\bibitem{folded}
  G.~Burdman, Z.~Chacko, H.~S.~Goh and R.~Harnik,
  JHEP {\bf 0702}, 009 (2007)
  [arXiv:hep-ph/0609152].

  
\bibitem{bj}
  L.~B.~Okun,
  JETP Lett.\  {\bf 31}, 144 (1980)
  [Pisma Zh.\ Eksp.\ Teor.\ Fiz.\  {\bf 31}, 156 (1979)];
  L.~B.~Okun,
  Nucl.\ Phys.\  B {\bf 173}, 1 (1980);
  J.~D.~Bjorken, (1979),
SLAC-PUB-2372;  
  S.~Gupta and H.~R.~Quinn,
  Phys.\ Rev.\  D {\bf 25}, 838 (1982).

\bibitem{quirks}
  J.~Kang and M.~A.~Luty,
  arXiv:0805.4642 [hep-ph].

\bibitem{Jacoby:2007nw}
  C.~Jacoby and S.~Nussinov,
  arXiv:0712.2681 [hep-ph];
  K.~Cheung, W.~Y.~Keung and T.~C.~Yuan,
  arXiv:0810.1524 [hep-ph].
  
\bibitem{squirk-ann}
  G.~Burdman, Z.~Chacko, H.~S.~Goh, R.~Harnik and C.~A.~Krenke,
  arXiv:0805.4667 [hep-ph].
  
  
\bibitem{pdg}
  W.~M.~Yao {\it et al.}  [Particle Data Group],
  J.\ Phys.\ G {\bf 33}, 1 (2006);
For a pedagogical review see {\tt http://www.physics.ucdavis.edu/~conway/talks/TASI/Conway-TASI-2.pdf}

\bibitem{atlfast}
  E. Richter-Was, D. Froidevaux and L. Poggioli, 
  ATLAS Note ATL-PHYS-98-131.

\bibitem{pgs}
{\tt  http://www.physics.ucdavis.edu/~conway/research/software/pgs/pgs4-general.htm}

\bibitem{sven}
see {\tt http://www.mppmu.mpg.de/~menke/}


\bibitem{Agostinelli:2002hh}
  S.~Agostinelli {\it et al.}  [GEANT4 Collaboration],
  Nucl.\ Instrum.\ Meth.\  A {\bf 506}, 250 (2003).

\bibitem{Sjostrand:2006za}
  T.~Sjostrand, S.~Mrenna and P.~Skands,
  JHEP {\bf 0605}, 026 (2006)
  [arXiv:hep-ph/0603175].


\bibitem{lattice}
  M.~J.~Teper,
  arXiv:hep-th/9812187.

\bibitem{BJ-fragmentation}
  J.~D.~Bjorken,
  Phys.\ Rev.\  D {\bf 17}, 171 (1978).

\bibitem{jackson}
J. D. Jackson, ``Classical Electrodynamics'', John Wiley, N.Y., 1998, Chapter 14. 

\bibitem{unparticles}
  H.~Georgi,
  Phys.\ Rev.\ Lett.\  {\bf 98}, 221601 (2007)
  [arXiv:hep-ph/0703260].

\bibitem{matt}
  M.~J.~Strassler,
  arXiv:0801.0629 [hep-ph].

\bibitem{feng}
  J.~L.~Feng, A.~Rajaraman and H.~Tu,
  arXiv:0801.1534 [hep-ph].

\bibitem{higgsportal}
  B.~Patt and F.~Wilczek,
  arXiv:hep-ph/0605188;
  Z.~Chacko, H.~S.~Goh and R.~Harnik,
  Phys.\ Rev.\ Lett.\  {\bf 96}, 231802 (2006)
  [arXiv:hep-ph/0506256];
  P.~J.~Fox, A.~Rajaraman and Y.~Shirman,
  Phys.\ Rev.\  D {\bf 76}, 075004 (2007)
  [arXiv:0705.3092 [hep-ph]].
  
\bibitem{MRST}
  A.~D.~Martin, W.~J.~Stirling, R.~S.~Thorne and G.~Watt,
  Phys.\ Lett.\  B {\bf 652}, 292 (2007)
  [arXiv:0706.0459 [hep-ph]].

\bibitem{TerningUnHiggs}
  D.~Stancato and J.~Terning,
  arXiv:0807.3961 [hep-ph].

\end{thebibliography}
\end{document}